\documentclass{IEEEtran4PSCC}

\usepackage{graphicx}
\usepackage[cmex10]{amsmath}
\usepackage{xfrac}
\usepackage{amsfonts}
\usepackage{enumitem}
\usepackage{epstopdf}
\usepackage{subfigure}
\usepackage{tikz}
\usetikzlibrary{shapes,arrows}
\usepackage{verbatim}
\usepackage{epstopdf}
\DeclareGraphicsExtensions{.eps}

\newcommand{\R}{\mathbb{R}}
\newcommand\m[1]{\begin{bmatrix}#1\end{bmatrix}}
\newcommand{\any}{\text{ $\forall$ }}
\newcommand{\x}{\mathbf{x}}
\newcommand{\A}{\mathbf{A}}
\newcommand{\Bw}{\mathbf{B_w}}
\newcommand{\Bu}{\mathbf{B_u}}
\newcommand{\C}{\mathbf{C}}
\newcommand{\Q}{\mathbf{Q}}
\newcommand{\z}{\mathbf{z}}
\renewcommand{\P}{\mathbf{P}}
\newcommand{\K}{\mathbf{K}}
\renewcommand{\S}{\mathbf{S}}
\newcommand{\W}{\mathbf{W}}
\newcommand{\e}{\mathbf{e}}
\renewcommand{\L}{\mathbf{L}}

\DeclareMathOperator*{\argmin}{arg\,min}

\makeatletter
\let\old@ps@headings\ps@headings
\let\old@ps@IEEEtitlepagestyle\ps@IEEEtitlepagestyle
\def\psccfooter#1{%
    \def\ps@headings{%
        \old@ps@headings%
        \def\@oddfoot{\strut\hfill#1\hfill\strut}%
        \def\@evenfoot{\strut\hfill#1\hfill\strut}%
    }%
    \def\ps@IEEEtitlepagestyle{%
        \old@ps@IEEEtitlepagestyle%
        \def\@oddfoot{\strut\hfill#1\hfill\strut}%
        \def\@evenfoot{\strut\hfill#1\hfill\strut}%
    }%
    \ps@headings%
}
\makeatother

\psccfooter{%
        \parbox{\textwidth}{\hrulefill \\ \small{21st Power Systems Computation Conference} \hfill \begin{minipage}{0.2\textwidth}\centering \vspace*{4pt} \includegraphics[scale=0.06]{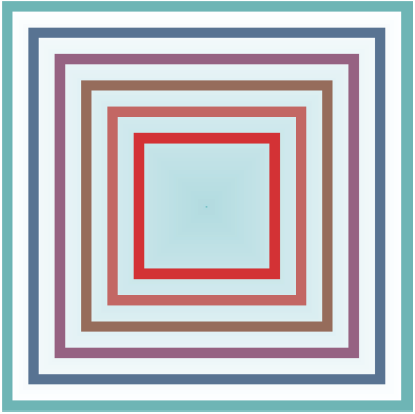}\\\small{PSCC 2020} \end{minipage} \hfill \small{Porto, Portugal --- June 29 -- July 3, 2020}}%
}

\begin{document}

\title{Optimal $L_\infty$ Frequency Control in Microgrids Considering Actuator Saturation}

\author{
\IEEEauthorblockN{Daniel Tabas and Baosen Zhang}
\IEEEauthorblockA{Department of Electrical and Computer Engineering\\
University of Washington\\
Seattle, WA 98195\\
\{dtabas, zhangbao\}@uw.edu}
}

\maketitle

\begin{abstract}
Inverter-connected resources can improve transient stability in low-inertia grids by injecting active power to minimize system frequency deviations following disturbances. In practice, most generation and load disturbances are step changes and the engineering figure-of-merit is often the peak overshoot in frequency resulting from these step disturbances. In addition, the inverter-connected resources tend to saturate much more easily than conventional synchronous machines. However, despite these challenges, standard controller designs must deal with averaged quantities through $H_2$ or $H_\infty$ norms and must account for saturation in ad hoc manners. In this paper, we address these challenges by explicitly considering $L_\infty$ control with saturation using a linear matrix inequality-based approach. We show that this approach leads to significant improvements in stability performance. 
\end{abstract}


\thanksto{\noindent
The authors are partially supported by the NSF Grant ECCS-1930605 and the University of Washington Clean Energy Institute.}
\section{Introduction}

As inverter-connected devices such as wind turbines, PV arrays, and batteries displace conventional synchronous generators, they will lead to a reduction in mechanical inertia in the system. At the same time, rising levels of variable renewable energy will create larger disturbances on the grid. Thus renewable energy resources place frequency stability under stress from two different directions. A power system that is unable to maintain operation close to nominal frequency in the face of high renewable penetration will be more prone to loss of synchronicity and subsequent loss of generation if new control strategies are not used.

Inverters bring unique advantages to the control of the grid and can offer performance characteristics that are more favorable than the conventional inertia of synchronous machines. Inverter-connected energy resources such as battery energy storage systems can provide the flexibility needed to absorb fluctuations in net demand. Changing the power tracking point of inverter-connected wind turbines and PV arrays can contribute additional flexibility. These systems are advantageous because the power inverters which serve as the interface between the generation and transmission systems can be controlled on much faster timescales than those on which grid electromechanical transients occur.

\begin{figure}
    \centering
    \tikzstyle{block} = [draw, rectangle, 
    minimum height=3em, minimum width=3em]
\tikzstyle{sum} = [draw, circle, node distance=1cm]
\tikzstyle{input} = [coordinate]
\tikzstyle{output} = [coordinate]
\tikzstyle{pinstyle} = [pin edge={to-,thin,black}]

% The block diagram code is probably more verbose than necessary
\begin{tikzpicture}[auto, node distance=1.75cm,>=latex']
    % We start by placing the blocks
    \node [input, name=input] {};
    \node [sum, right of=input] (sum) {};
    \node [block, right of=sum] (controller) {Controller};
    \node [block, right of=controller] (saturation) {sat($\cdot$)};
    \node [block, right of=saturation, pin={[pinstyle]above:Disturbance}] (system) {Plant};
    % We draw an edge between the controller and system block to 
    % calculate the coordinate u. We need it to place the measurement block. 
    \draw [->] (controller) -- node[name=q] {$q$} (saturation);
    \draw [->] (saturation) -- node[name=u] {$u$} (system);
    \node [output, right of=system] (output) {};
    \node [block, below of=saturation] (measurements) {Measurements};

    % Once the nodes are placed, connecting them is easy. 
    \draw [draw,->] (input) -- node {$r$} (sum);
    \draw [->] (sum) -- node {$e$} (controller);
    \draw [->] (system) -- node [name=y] {$y$}(output);
    \draw [->] (y) |- (measurements);
    \draw [->] (measurements) -| node[pos=0.99] {$-$} 
        node [near end] {$\hat{y}$} (sum);
\end{tikzpicture}
    \caption{Block diagram of a system with actuator saturation. In our setting, $r$ is the reference frequency to be tracked under uncertain disturbances, where the actuators have saturation constraints.}
    \label{fig:blockdiagram}
\end{figure}
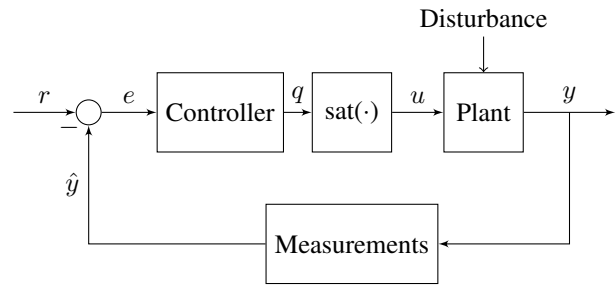

In this paper, we present a method for controller synthesis considering two complexities encountered in power system frequency regulation. First, the proposed synthesis accounts for constraints on controller output corresponding to power limits on inverters. Since inverter power ratings are typically several orders of magnitude smaller than those of synchronous machines \cite{Kroposki2017}, it can be expected that inverters providing frequency support will often be operating at or near their power limit. Conventional methods of frequency support such as droop control are designed without accounting for saturation, and inverters mimicking them (e.g., virtual synchronous machines) can suffer in real systems when saturation is unavoidable. Figure \ref{fig:blockdiagram} illustrates the configuration of a simple feedback system with actuator saturation.

Second, the proposed synthesis minimizes the $L_\infty$-norm of the closed-loop system, since the ability of a power system to withstand disturbances is often evaluated in terms of maximum overshoot rather than the $H_2$ or $H_\infty$ norms commonly adopted in control literature ~\cite{Kroposki2017}. It is the peak overshoot that determines how close the full nonlinear system gets to its stability limit.

The method of invariant ellipsoids arises as a natural way to deal with the joint complexities of actuator saturation and $L_\infty$ norm minimization because it has been successfully applied to each of these problems individually. The convexity of the linear matrix inequalities defining these invariant ellipsoids allows the LMIs to be manipulated and combined to form semidefinite programs generating new invariant ellipsoids that address the two challenges simultaneously. Saturation is treated extensively using set invariance in \cite{Boyd1994} and \cite{Hu2001}. Ref. \cite{Nguyen1999} develops sufficient conditions for guaranteed $L_\infty$ performance in systems with saturation, while \cite{Nguyen1999a} provides $L_2$-gain results for systems with actuator rate constraints. Meanwhile, optimal $L_\infty$ controller design is addressed in \cite{Abedor1996}.

Some of these results have been applied to power system regulation problems. Guarantees on damping ratio and settling time were developed in \cite{soliman2015}. Disturbance rejection guarantees using invariant ellipsoids were presented in \cite{Soliman2018}, while \cite{Xin2007} and \cite{Xin2008} used invariant ellipsoids to estimate stability regions. However, none of these papers deals explicitly with minimizing the peak overshoot of the system.

The current paper presents an LMI-based technique for optimizing an inverter controller with saturation nonlinearities, with the goal of minimizing the maximum frequency deviation of the system following a disturbance. The contributions are twofold. First, we combine results from \cite{Nguyen1999} and \cite{Abedor1996} to present a unified approach to optimal $L_\infty$ controller synthesis in systems with actuator saturation. Second, we demonstrate a first application of these controller synthesis results to power system frequency regulation. The performance guarantees presented in this paper only assume that the disturbance is bounded in $L_\infty$ norm, and can otherwise be an arbitrary signal.

The paper is organized as follows. Section \ref{sec:model} discusses the power system model under consideration. Section \ref{sec:linf} presents the mathematical background underpinning section \ref{sec:ctrl}, control design. Section \ref{sec:sims} presents the results of the controller design applied to the power system model discussed in section \ref{sec:model}.

\section{Model} \label{sec:model}
We consider a simplified single-area system consisting of an aggregate generator and turbine governor, an inverter-connected energy resource, and a load. The frequency dynamics of the generator are given by the linearized swing equation and the linearized turbine governor equation:
\begin{subequations} \label{eqn:system} 
\begin{align}
    M \dot{\Delta \omega} + D \Delta \omega = \Delta P_\text{M} - \Delta P_\text{L} + \Delta P_\text{I}\\
    \dot{\Delta P_\text{M}} = - \frac{k}{\rho} \Delta \omega - k \Delta P_\text{M},
\end{align}
\end{subequations}
where $\Delta \omega$ is the change in generator frequency following a disturbance, $M$ and $D$ are the inertia and damping constants of the simplified generator model, $\Delta P_\text{M}$ is the change in power output from the prime mover, $\Delta P_\text{M}$ is the change in real power demand, $\Delta P_\text{I}$ is the real power injection from the inverter, and $\rho$ is the speed-droop coefficient of the turbine governor \cite{Kundur1994}. Inverter dynamics are not modeled. 

We take $\Delta P_\text{L}$ as the external disturbance into the system and assume it is has bounded infinity norm. Then without loss of generality, assume $||\Delta P_\text{L}||_\infty=1$. The goal is to design a controller for $\Delta P_\text{I}$ to minimize the impact of this disturbance on the system, subject to inverter saturation constraints. 

The inverter power injection is modeled as $\Delta P_\text{I}(t)=\text{sat}(q)$,  where $q$ is a control input and the saturation function is given by $\text{sat}(z ) = \text{sgn}(z )\min(|z |,1)$. In general, the form of the optimal controller is not known. We propose a control law of the form $q = -\K \x, \Delta P_\text{I} = \text{sat}(q)$ where $\K $ is a linear feedback gain matrix to be optimized and $\x $ is the state of the system, given by $\m{\Delta \omega & \Delta P_\text{M}}^T$. In the proceeding sections, we consider both full-state feedback and output feedback in which only $\Delta \omega$ is measured.

\section{Performance Metrics} \label{sec:linf}

\subsection{$L_\infty$-norms}

    We define the $L_\infty$-norm of a signal $\z (t) \in \R^n$ as
    \begin{gather}
        \| \z (t)\|_\infty = \sup_{t} \| \z (t)\|_2.
    \end{gather}
    
    \noindent Now consider the open-loop system $H$:
            \begin{align}
                \dot{\x } &= \A \x  + \Bw w \nonumber \\
                y &= \C \x, \label{eq:sys}
            \end{align}
    where $w$ is an exogenous disturbance of bounded infinity norm. Throughout the paper, we assume that $(\A,\Bw)$ is controllable and $(\A,\C)$ is observable.   Without loss of generality, we can normalize $w$ such that $\|w\|_\infty \leq 1$. The disturbance can represent changes in net demand or unexpected changes in generation. The reachable set of the open-loop system initialized at the origin is defined as
            \begin{align}
                &\mathcal{R}_{OL} = \bigg\{\int_0^t e^{(t-\tau)\A }\Bw w(\tau)d \tau : t \geq 0, \|w\|_\infty \leq 1 \bigg\} \label{eq:reachable}
            \end{align}
    
    The $L_\infty$-norm of the system \eqref{eq:sys} is the largest possible measured output:
    \begin{gather}
        \sup_{\x  \in \mathcal{R}_{OL}} \|\C \x \|_2
    \end{gather}
    
    In the frequency regulation problem with states $\x = \m{\Delta \omega & \Delta P_m}^T,$ the quantity of importance is $\Delta \omega$, and this is also typically the observable quantity. The $L_\infty$ norm of the system corresponds to the maximum frequency deviation possible given norm-bounded disturbances.

\subsection{$L_\infty$-norm overbound: the $*$-norm}
        Ideally, we would want to characterize the reachable set, but it turns out that computing it directly is difficult.  In a system with $m$ saturating elements, there are $3^m$ regions with different dynamics, each corresponding to a different combination of saturating element states. Instead of dealing explicitly with these saturation nonlinearities, we approximate the reachable set using invariant ellipsoids, from which the $*$-norm, an upper bound on the $L_\infty$ norm, can be computed. First, consider the open-loop system. Using Lyapunov stability, $\mathcal{R}_{OL}$ is contained within the invariant ellipsoid $\varepsilon_1 = \{\x  \mid \x ^T\Q ^{-1}\x  \leq 1 \}$, for any $\Q \succeq 0$ where an $\alpha>0$ can be found such that $\Q$ satisfies~\cite{Abedor1996}
        \begin{gather}
            \m{\A \Q  + \Q \A^T + \alpha \Q  & \Bw  \\ \Bw ^T & -\alpha \mathbf{I}} \preceq 0 \label{eq:OL_Lyap}.
        \end{gather}

        To understand where LMI \eqref{eq:OL_Lyap} comes from, consider a quadratic Lyapunov function $V(\x(t)) = \x(t)^T \Q^{-1} \x(t)$ and recall that Lyapunov stability inside the ellipsoid $\varepsilon_1$ for a system with norm-bounded disturbances requires that $\dot{V}(\x) \leq 0$ outside $\varepsilon_1$ whenever $\|w\| \leq 1$ (argument $t$ omitted for brevity). These conditions condense down to the requirement that $\dot{V}(\x) \leq 0$ whenever $[w^Tw - \x^T \Q^{-1} \x] \leq 0$. Applying the S-lemma, an equivalent condition is $\dot{V}(\x) \leq \alpha [w^Tw - \x^T \Q^{-1} \x]$ for some $\alpha \geq 0.$ Next, use the chain rule to evaluate $\sfrac{\partial}{\partial t} V(\x)$, make the substitution $\dot{\x} = \A \x + \Bw w$, and rewrite the inequality as a quadratic form in $\m{\x &w }^T.$ Correspondence between nonnegativity of a quadratic form and positive semidefiniteness of the associated matrix yields LMI \eqref{eq:OL_Lyap}. Lyapunov stability dictates that the system initialized at the origin cannot leave $\varepsilon_1$ when $\|w\|_\infty \leq 1$. Therefore, $\varepsilon_1$ contains or is the reachable set.
        
        Similarly to the way in which we defined $\|H\|_{L_\infty}$ as the largest possible measured output within the reachable set, we can use $\varepsilon_1$ to define $N_\alpha$, the largest possible measured value within $\varepsilon_1$. Since $\varepsilon_1$ contains the reachable set, it can be used to define an upper bound on $\|H\|_{L_\infty}$. For a fixed value of $\alpha$, $N_\alpha$ is given by
        \begin{gather}
            N_\alpha = \sup_{\x  \in \varepsilon_1} \|\C \x \|_2 \label{eq:Na}
        \end{gather}
        where $\varepsilon_1$ is defined by a positive semi-definite $\Q$ satisfying \eqref{eq:OL_Lyap} \cite{Abedor1996}. For the frequency regulation problem, the measured output $\C \x$ is $\Delta \omega.$ Since we are free to choose $\alpha$ within the constraints of \eqref{eq:OL_Lyap}, we can minimize our overestimate of the $L_\infty$-norm by, at worst, searching over $\alpha$ on an interval \cite{Abedor1996}. The minimum value resulting from this line search, the $*$-norm, is given by
        \begin{gather}\begin{split}
            \|H\|_* = \min_{\alpha \geq 0, \Q  \succeq 0} N_\alpha \\
            \text{subject to } \eqref{eq:OL_Lyap}. \label{eq:starnorm}
        \end{split}\end{gather}
        
        The corresponding ellipsoid $\{\x  \mid \x ^T\Q ^{-1}\x  \leq 1 \}$ provides the best (smallest) ellipsoidal over-approximation of the reachable set. By substituting the dual of the square of \eqref{eq:Na} into \eqref{eq:starnorm}, we obtain an equivalent expression for the $*$-norm
        \begin{subequations}\label{eq:starnorm_final}
        \begin{gather}
            \|H\|_*^2 = \min_{\alpha,\lambda,\Q } \lambda\\
            \text{s.t. } \alpha, \lambda \geq 0; \Q  \succeq 0;\\
            % \eqref{eq:OL_Lyap};
                   \m{\A \Q  + \Q \A^T + \alpha \Q  & \Bw  \\ \Bw ^T & -\alpha \mathbf{I}} \preceq 0 \label{eq:OL_Lyap_repeat} \\
    % \tag{\ref{eq:starnorm_final}} \\
            \m{\lambda & \C \Q  \\ \Q \C ^T & \Q } \succeq 0 \label{eq:starnormLMI}.
        \end{gather}
        \end{subequations}
        
        The LMIs in \eqref{eq:OL_Lyap_repeat}  (or \eqref{eq:OL_Lyap}) and \eqref{eq:starnormLMI} are convex in $(\Q, \lambda)$ and therefore provide an efficient way of computing $N_\alpha$ for a fixed $\alpha$. The $*$-norm is powerful because it provides a computationally efficient way of approximating a system's $L_\infty$ norm. Next, we utilize the tractability of the $*$-norm to present a method for optimal $L_\infty$ controller synthesis.
\section{Control Design} \label{sec:ctrl}

We present results based on \cite{Nguyen1999} and \cite{Abedor1996} for synthesis of full-state and output feedback controllers optimized for $L_\infty$ performance considering hard constraints on control output. The system equation in this case becomes:
\begin{align}
    \dot{\x } &= \A \x  + \Bw w +\Bu u \nonumber \\
    y &= \C \x, \label{eq:sys_input}
\end{align}
where $\Bu$ is the control matrix. In our case, $u$ is the active power injection of the inverters, and $\Bu=[1 \; 0]^T$. 

For each feedback paradigm, we present the synthesis of a linear ``low-gain'' feedback controller and a nonlinear, saturating ``high-gain'' controller based on minimizing the $*$-norm of the closed loop system. The low-gain controller relies on a guarantee that the saturating element will never be activated within the reachable set. The high-gain controller utilizes a crucial fact from \cite{Nguyen1999} to guarantee that its performance will be at least as good as the performance of the optimal low-gain controller.

    \subsection{Full-state feedback}
    
    \subsubsection{Low-gain/linear controller}
    
        We first consider a linear feedback control law of the form $u = -\K\x$ given by 
        \begin{gather}
            u = -\frac{v}{2}\Bu^T\Q ^{-1} \x \label{eq:u}
        \end{gather} 
        for some $v > 0$ and $\Q \succeq 0$. This control law is sufficient for closed-loop stability since it results in convenient LMIs, but is not the only possible control law. Writing down \eqref{eq:OL_Lyap} for the closed loop system by substituting $\A-\Bu \K$ for $\A$ leads to a condition on $(\Q,v,\alpha)$ for closed-loop Lyapunov stability inside $\{\x \mid \x ^T\Q ^{-1}\x \leq 1\}$ given by 
        \begin{gather}
            \m{\A \Q  + \Q \A ^T -v\Bu \Bu^T+ \alpha \Q  & \Bw \\ \Bw^T & -\alpha} \preceq 0. \label{eq:FS_CL_Lyap}
        \end{gather}
        
        It immediately follows that the $*$-norm-optimal controller for a linear system can be obtained by replacing \eqref{eq:OL_Lyap_repeat} with \eqref{eq:FS_CL_Lyap} and solving \eqref{eq:starnorm_final}. A sufficient way to ensure that this approach will work in systems with saturation nonlinearities is to ensure that the saturation elements are never activated. Specifically, this means enforcing $\|u\| \leq u_\text{max}$ within $\{\x \mid \x ^T\Q ^{-1}\x \leq 1\}$ when designing the linear feedback controller gains.
        In order to translate this into a constraint on $\Q$ and $v,$ we start by rewriting $\x^T \Q^{-1} \x \leq 1 \implies \|u\|^2 \leq u_\text{max}^2$ as $\frac{u^Tu}{u_\text{max}} \leq \x^T \Q^{-1} \x$. Substituting for $u$ using \eqref{eq:u} and rearranging using the Schur complement leads to the additional LMI in $(\Q,v)$ given by \cite{Nguyen1999}
        \begin{gather}
            \m{4\Q  & v\Bu \\ v\Bu^T & u_\text{max}^2} \succ 0. \label{eq:FS_ubound}
        \end{gather}
        
        However, due to \eqref{eq:FS_ubound}, full control effort is only realized at the boundary of the invariant ellipsoid. To see why, consider a point $\x^{(1)}$ on the interior of $\{\x \mid \x ^T\Q ^{-1}\x \leq 1\}$ where $\|\K \x^{(1)}\| = u_\text{max}.$ Then scaling $\x^{(1)}$ by $\mu > 1$ to extend it to the boundary of the ellipsoid would result in a control effort of $\|\mu \K \x^{(1)}\| > u_\text{max},$ which contradicts \eqref{eq:FS_ubound}. It is reasonable to expect superior performance if the full control effort can be utilized on the interior as well as the boundary of the ellipsoid. A mechanism for achieving this is discussed in the next section.
        
        \subsubsection{High-gain/nonlinear controller}
        In order to utilize more control capacity more of the time without sacrificing performance, we start with the controller from the previous section, scale it by $\delta > 1$, and saturate the result at $\pm u_\text{max}$, giving a controller of the form $u(t) = -\text{sat}(\delta \frac{v}{2}\Bu^T\Q ^{-1} \x(t))$, with $\Q \succeq 0$ satisfying \eqref{eq:FS_CL_Lyap} and \eqref{eq:FS_ubound} \cite{Nguyen1999}. The performance guarantee for this controller relies on a crucial fact: for any $(\Q,v,\alpha)$ satisfying \eqref{eq:FS_CL_Lyap} and \eqref{eq:FS_ubound}, the closed-loop system is Lyapunov stable within the ellipsoid $\{\x \mid \x ^T\Q ^{-1}\x \leq 1\}$ for both the low-gain and high-gain controllers. For a proof, see \cite{Nguyen1999}. This means that the $*$-norm of the closed-loop system under the high-gain controller is no larger than that of its low-gain counterpart. In fact, we can expect better performance from the high-gain controller due to increased control expenditure. As $\delta \rightarrow \infty$, the behavior of the high-gain controller approaches bang-bang or on/off control. 
        
        To summarize, the full-state feedback synthesis problem is to pick a $\Q$ and $v$ such that 
        \begin{gather*}
            \Q , v = \argmin_{\alpha,\lambda,v,\Q } \lambda \\
            \text{s.t. } \lambda \geq 0; \alpha, v > 0, \Q  \succ 0; \\
            \eqref{eq:starnormLMI}, \eqref{eq:FS_CL_Lyap}, \eqref{eq:FS_ubound}.
        \end{gather*}
        
        These results, while useful, do not reflect the true nature of the single-area frequency regulation problem because of the assumption that all states are perfectly measured. In the next section, we present results for saturating control of output feedback systems.
        
    \subsection{Observer-based output feedback}
    We proceed in parallel to the full-state feedback results. For output feedback, we consider the system
        \begin{align*}
            \dot{\x} &= \A \x + \Bw w + \Bu u\\
            \dot{\hat{\x}} &= \A \hat{\x} + \Bu u + \L(y -  \C_2 \hat{\x})\\
            y &=  \C \x
        \end{align*}
        where $\x$ is the state, $\hat{\x}$ is the measurement or estimate, and $y$ is the measured output. The synthesis proceeds as follows: first, a full state feedback controller is designed. Then, using the full state feedback control gain, the observer gain $\L$ is designed using two LMIs that are analogous to \eqref{eq:FS_CL_Lyap} and \eqref{eq:FS_ubound} \cite{Nguyen1999}.
        
        \subsubsection{Low-gain/linear controller}
        Define the error signal $\e(t) = \x(t) - \hat{\x}(t)$. Let $\L = \S^{-1} \W$ for some $\S \succ 0$, and $\P = \Q^{-1}$ as computed in the full state feedback controller design. Stacking the plant and error states, an invariant ellipsoid is given by $$\bigg \{\x,\e \in \R^n \mid \m{\x& \e} \m{\P & 0\\ 0 & \S} \m{\x \\ \e} \leq 1 \bigg \}$$ where $\S$ and $\W$ satisfy
        \begin{flalign}
        &\left[\begin{matrix}
            \P \A  + \A ^T\P  - v\P \Bu \Bu^T\P  + \alpha \P \\
            \frac{v}{2}\P \Bu \Bu^T\P  \\
            \Bw^T\P  
        \end{matrix}\right.\nonumber &
        \end{flalign}
        \begin{align}
        \left.\begin{matrix}
            \frac{v}{2}\P \Bu  \Bu^T\P  & \P \Bw\\
            \S \A  + \A ^T\S - \W  \C _2 -  \C _2^T\W ^T + a\S & \S \Bw\\
            \Bw^T\S & -\alpha
        \end{matrix}\right]
        \preceq 0. \label{eq:OF_CL_Lyap}%\any \varphi \in \{1,\delta\}
        \end{align}
         This LMI is derived from \eqref{eq:OL_Lyap} the same way that \eqref{eq:FS_CL_Lyap} is, and guarantees that the observer is stable.
        
        The LMI guaranteeing that $\|u\|$ stays below the saturation limit $u_\text{max}$ is given by \cite{Nguyen1999}
        \begin{gather}
            \m{\P  & 0& -\frac{v}{2}\P \Bu\\
                0 & \S & \frac{v}{2}\P \Bu\\
                -\frac{v}{2}\Bu^T\P  & \frac{v}{2}\Bu^T\P  & u_\text{max}^2} \succ 0. \label{eq:OF_ubound}
        \end{gather}
        This LMI is analogous to \eqref{eq:FS_ubound} and is derived in the same way using the augmented states.
        
        \subsubsection{High-gain/nonlinear controller}
        Following the intuition of the full-state feedback controller synthesis, the high-gain controller for output feedback is given by $ u(t) = - \text{sat}(\delta \frac{v}{2} \Bu^T \Q ^{-1} \hat{\x}(t))$, where $\Q \succeq 0$ satisfies \eqref{eq:FS_CL_Lyap} and \eqref{eq:FS_ubound}, and $\delta > 1$ \cite{Nguyen1999}.
        
        The action of the high-gain saturating controller must be accounted for in the condition for observer stability \eqref{eq:OF_CL_Lyap}. Note that since the low-gain controller design guarantees $\|\frac{v}{2} \Bu^T \Q ^{-1}\x\| \leq u_\text{max},$ the action of the high-gain controller is equivalent to some scaling of the low-gain controller by a factor $r$ which varies from 1 when both the low- and high-gain control magnitudes saturate at $u_\text{max}$, to $\delta$ when neither does. Representing the high-gain controller as the low-gain controller scaled by $r$ leads to 
        
        \begin{flalign}
        &\left[\begin{matrix}
            \P \A  + \A ^T\P  - rv\P \Bu \Bu^T\P  + \alpha \P \\
            r\frac{v}{2}\P \Bu \Bu^T\P  \\
            \Bw^T\P  
        \end{matrix}\right.\nonumber &
        \end{flalign}
        \begin{align}
        \left.\begin{matrix}
            r\frac{v}{2}\P \Bu  \Bu^T\P  & \P \Bw\\
            \S \A  + \A ^T\S - \W  \C _2 -  \C _2^T\W ^T + a\S & \S \Bw\\
            \Bw^T\S & -\alpha
        \end{matrix}\right]
        \preceq 0 \label{eq:OF_CL_Lyap_hg}
        \end{align}
        which must hold for any $r \in [1,\delta]$, but since the LMI is convex in $r$, this is guaranteed if it holds for $r \in \{1,\delta\}$ \cite{Nguyen1999}.
        
        \subsubsection{Output feedback controller synthesis}
        We now discuss several nuances related to the synthesis of an output feedback controller as described above. When selecting $(\Q,v,\alpha)$ for controller design, we additionally enforce LMI \eqref{eq:OL_Lyap} in order to ensure that the chosen combination of ($\Q ,v,\alpha$) admits a feasible solution to the observer design problem. The sufficiency of this condition was verified experimentally. When selecting $(\S,\W )$ from the feasible region of the observer design problem, we choose values that minimize the $*$-norm of the measurement error. This results in the introduction of another dual variable $\theta \geq 0$ and an additional LMI analogous to \eqref{eq:starnormLMI}:
        \begin{gather}
            \m{\theta & \C \\  \C ^T & \S} \succeq 0 \label{errorstarnormLMI}
        \end{gather}
        
        Therefore, we can summarize the design of the high-gain observer-based output feedback controller in the following way:\\
        \textbf{Controller design:}
            \begin{gather*}
                \Q , v = \argmin_{\alpha,\lambda,v,\Q } \lambda \\
                \text{s.t. } \alpha, \lambda, v \geq 0; \Q  \succeq 0; \\
                \eqref{eq:OL_Lyap}, \eqref{eq:starnormLMI}, \eqref{eq:FS_CL_Lyap}, \eqref{eq:FS_ubound}\\
            \end{gather*}
        \textbf{Observer design:}
            \begin{gather*}
                \S ,\W  = \argmin_{\theta, \S , \W } \theta\\
                \text{s.t. } \theta \geq 0, \S \succ 0;\\
                \eqref{eq:OF_ubound},\eqref{errorstarnormLMI}\\
                \eqref{eq:OF_CL_Lyap_hg}, \any r \in \{1,\delta\}
            \end{gather*}
            
    The controller design problem is convex in $(\Q,v,\lambda)$ and consists of a line search over $\alpha$. Meanwhile, the observer design problem is convex in $(\S,\W)$. Since these are linear matrix inequalities, they can be solved using standard convex optimization techniques. 
    % Due to the linear matrix inequality techniques employed throughout this paper, the computational complexity of the problem remains low.
\section{Simulation Results} \label{sec:sims}

\subsection{Full state feedback}

We simulated the full-state feedback system with the following parameters: $M = 2,
D = 0.6,
\rho = 0.05,
k = 5 ,
w_{max} = 0.1 \text{ p.u.}, $ and $
u_{max} = 0.05 \text{ p.u.}.$ The resulting $*$-norm-optimal controller is given by $\K = \m{2.89    & 0.0808} .$

\begin{figure}[ht]
    \centering
    \includegraphics[width = 2.5in]{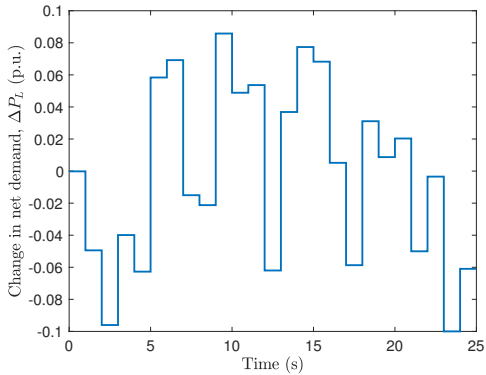}
    \caption{Disturbance profile.}
    \label{subfig:disturbances}
\end{figure}

\begin{figure}[ht]
    \centering
    \subfigure[]{\includegraphics[width=2.5in]{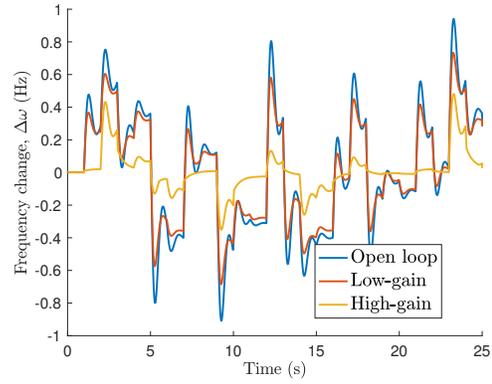}%
    \label{subfig:FS_TD}}
    %\hfil

    \subfigure[]{\includegraphics[width = 2.5in]{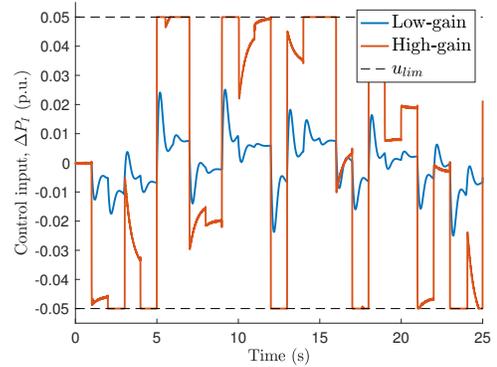}
    \label{subfig:FS_Control_inputs}}
    \caption{(a) Time-domain response of the open loop system and the closed-loop full-state feedback system with low- and high-gain controllers. (b) Control inputs for the low- and high-gain controllers.}
    \label{fig:FS1}
\end{figure}

To simulate the system, we applied a signal $w(t)$ comprised of random step changes satisfying $\|w(t)\|_\infty \leq w_{max},$ shown in Fig. \ref{subfig:disturbances}. The responses of the open-loop system and the closed-loop system with low-gain controller and high-gain controller with $\delta = 100$ are shown in Fig. \ref{subfig:FS_TD}. The low-gain controller mitigates overshoot to an extent, but the high-gain controller performs significantly better for reasons that are evident in Fig. \ref{subfig:FS_Control_inputs}: the controller spends more time at maximum output.

\begin{figure}[ht]
    \centering
    \subfigure[]{\includegraphics[width=2.5in]{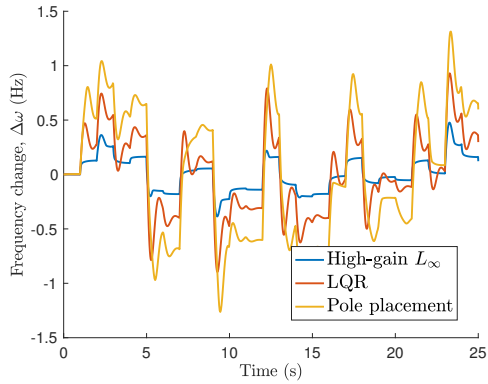}%
    \label{subfig:FS_Comparison}}
    %\hfil

    \subfigure[]{\includegraphics[width=2.5in]{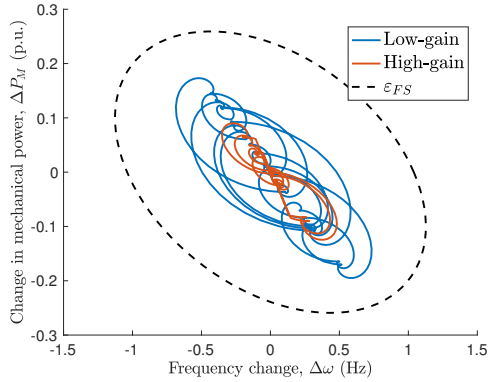}%
    \label{subfig:FS_PP}}
    \caption{(a) Time-domain comparison of $\Delta \omega$ trajectories for full-state feedback LQR, pole placement, and high-gain (with $\delta = 10$) $L_\infty$ controller designs. (b) Invariant ellipsoid for full-state feedback and two phase plane trajectories of the full-state feedback system corresponding to the low-gain controller and the high-gain controller with $\delta = 100$. }
    \label{fig:FS}
\end{figure}

We used LQR and pole placement controller designs for baseline comparisons to the proposed controller. The feedback matrices were $K = \m{0.138 &   0.0045}$ for LQR, and $K = \m{1.70  &  0.480}$ for pole placement. A gain of $\delta = 10$ was used for the high-gain $L_\infty$ controller. Fig. \ref{subfig:FS_Comparison} compares the performance of the proposed controller to that of the baseline controllers subjected to saturation. The proposed controller performs better than conventional methods and comes with additional guarantees about worst-case performance.

We simulated a series of disturbances rather than a single disturbance in order to ``probe'' the reachable set. Fig. \ref{subfig:FS_PP} juxtaposes the $*$-norm-optimal invariant ellipsoid and the states that might be occupied by the system when subjected to an $L_\infty$-norm-bound disturbance. It is important to note that the phase plane trajectories presented in Fig. \ref{subfig:FS_PP} do not fill the reachable set. Rather, it is hoped that they cover a large portion of the reachable set in order to illustrate the accuracy of the invariant ellipsoid as an approximation of the reachable set. While Fig. \ref{subfig:FS_PP} does not show that $\varepsilon_{FS}$ is a tight upper bound on the reachable set, it does show that the $*$-norm of the system and the $L_\infty$-norm of the system's response to the disturbances $w(t)$ in Fig. \ref{subfig:disturbances} are well within an order of magnitude. The two trajectories shown in Fig. \ref{subfig:FS_PP} correspond to the low-gain controller and the high-gain controller with $\delta = 100$. Fig. \ref{subfig:FS_Control_inputs} shows that at a gain of $\delta = 100$, the control behavior approaches bang-bang control, thus performance is not expected to increase significantly with higher gain.

\subsection{Output feedback}

\begin{figure}[ht]
    \centering
    \subfigure[]{\includegraphics[width=2.5in]{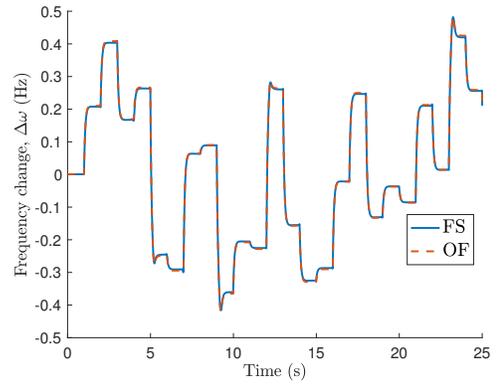}%
    \label{subfig:OF_TD}}
    %\hfil

    \subfigure[]{\includegraphics[width=2.5in]{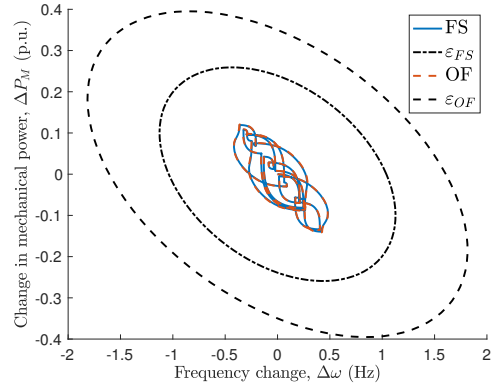}%
    \label{subfig:OF_PP}}
    \caption{(a) Time-domain comparison of $\Delta \omega$ trajectories for the full-state and output feedback high-gain controllers with $\delta = 10$. (b) Comparison of the invariant ellipsoids and phase plane trajectories of the full-state and output feedback systems for the same high-gain controller.}
    \label{fig:OF}
\end{figure}

The output feedback results in Fig. \ref{fig:OF} show near-perfect tracking with $*$-norm-optimal feedback gain $\K = \m{1.381 &    0.0491}$ and observer gain $\L = 10^3 \times \m{1.11 & -0.100}^T.$ The reason for the discrepancy in $\K$ between full-state and output feedback is the implementation of the additional LMI \eqref{eq:OL_Lyap} in the controller design as an ad-hoc way of guaranteeing observer feasibility. The simulation results show identical performance between the two controllers, but a substantial penalty in the $*$-norm guarantee for giving up $\Delta P_\text{M}$ measurements.

\section{Conclusion}

We have presented a self-contained approach to $L_\infty$-optimal feedback controller design in systems with actuator saturation. Such systems are of ever-increasing prevalence in power systems as power electronics-connected devices populate the grid. Simulation results show that for a simplified single-area power system model, the proposed controller substantially outperforms conventional controller designs such as LQR control and pole placement. Further, the results show that for the model used in this paper, the $*$-norm provides a reasonable, if not tight, overbound of the system's $L_\infty$ norm. Future work includes implementation in a multi-node system with multiple controllable devices.
\bibliographystyle{ieeetr}
\bibliography{PSCC2020}

\end{document}